\documentclass[11pt,letterpaper]{article}
\usepackage[margin=1in]{geometry}
\usepackage{amssymb,amsmath,amsthm}
\usepackage{graphicx}
\usepackage[colorlinks,citecolor=blue,linkcolor=blue,urlcolor=red,pagebackref]{hyperref}
\usepackage{subcaption}
\usepackage{booktabs}
\usepackage{color}
\usepackage{enumitem}
\usepackage{tikz}
\usepackage[noend]{algpseudocode}
\usepackage{framed}
\usepackage{thm-restate}
\usepackage{algorithm}
\usepackage{multirow}
\usepackage[T1]{fontenc}
\usepackage{mathtools,microtype}
\usepackage{xurl}

\newtheorem{theorem}{Theorem}

\newtheorem{corollary}[theorem]{Corollary}

\newtheorem{lemma}[theorem]{\bfseries{Lemma}}

\newtheorem{assumption}[theorem]{Assumption}

\newcommand{\eps}{\varepsilon}
\newcommand{\Ot}{\widetilde{O}}
\DeclareMathOperator{\poly}{poly}

\newbox\ProofSym
\setbox\ProofSym=\hbox{%
\unitlength=0.18ex%
\begin{picture}(10,10)
\put(0,0){\framebox(9,9){}}
\put(0,3){\framebox(6,6){}}
\end{picture}}

\title{A Note on Truly Subquadratic Time FPTAS for Knapsack}
\author{
  Lin Chen\thanks{chenlin198662@zju.edu.cn. Zhejiang University.}
  \and
  Jiayi Lian\thanks{jiayilian@zju.edu.cn. Zhejiang University.}
  \and
  Yuchen Mao\thanks{maoyc@zju.edu.cn. Zhejiang University.}
  \and
  Guochuan Zhang\thanks{zgc@zju.edu.cn. Zhejiang University.}
}
\date{}
\hypersetup{
  pdftitle={A Note on Truly Subquadratic Time FPTAS for Knapsack},
  pdfauthor={Lin Chen, Jiayi Lian, Yuchen Mao, and Guochuan Zhang}
}

\begin{document}
\maketitle

\begin{abstract}
We show that an $\Ot(m^\rho)$-time algorithm for integer $(\min,+)$-convolution,
where $1<\rho<2$, gives a randomized fully polynomial-time approximation
scheme for $0$--$1$ Knapsack running in
$\Ot(n+(1/\eps)^{1+\rho/2})$ time. The algorithm returns a feasible solution
whose profit is at least the optimum divided by $1+\eps$, with high
probability. We use the all-capacities proximity decomposition of Chen,
Lian, Mao, and Zhang. Its two quadratic computations are replaced by
blocked computation of requested prefix and suffix functions, and exact
computation within groups sharing one of Chan's profit grids. Thus the
approximation scheme retains half of the saving in the convolution
exponent. Wang's weighted-cache 3SUM bound, transferred to witnessed
integer convolution by Zhou, gives the
numerical bound $\Ot(n+(1/\eps)^{1.981})$. 
\end{abstract}

\section{Introduction}\label{sec:introduction}
We consider the classic Knapsack problem. Given a capacity $t$ and a set
$I$ of $n$ items, where item $i$ has weight $w_i$ and profit $p_i$, our
goal is to find a subset $I'\subseteq I$ that maximizes
$\sum_{i\in I'}p_i$ subject to $\sum_{i\in I'}w_i\le t$.
Throughout this paper, weights are positive integers, profits are
nonnegative integers, and the capacity is a nonnegative integer.
A fully polynomial-time approximation scheme (FPTAS), given
$\eps\in(0,1]$, returns a feasible solution within a factor of $1+\eps$
of the optimum in time polynomial in the input length and $1/\eps$.

Since the first FPTAS of Ibarra and Kim~\cite{IK75}, there has been a
long line of research on improving the dependence on $1/\eps$
\cite{Law79,KP04,Chan18,Jin19,DJM23}. Chen, Lian, Mao, and
Zhang~\cite{CLMZ25} and, independently, Mao~\cite{Mao24} obtained
$\Ot(n+(1/\eps)^2)$-time FPTASes. Table~\ref{tab:history} summarizes
some of these bounds. The reductions of Cygan, Mucha, W{\k{e}}grzycki,
and W{\l}odarczyk~\cite{CMWW19}
and K\"unnemann, Paturi, and Schneider~\cite{KPS17} rule out an
$O((n+1/\eps)^{2-\delta})$-time FPTAS for any fixed $\delta>0$
under the $(\min,+)$-convolution hypothesis.

\begin{table}[t]
\centering
\caption{Selected FPTASes for Knapsack. Polylogarithmic factors are
suppressed. A dagger indicates randomization; the last row assumes
integer convolution exponent $\rho$ as in Assumption~\ref{ass:convolution}.}
\label{tab:history}
\begin{tabular}{@{}ll@{}}
\toprule
Running time & Reference\\
\midrule
$\Ot(n+(1/\eps)^4)$ & Ibarra and Kim~\cite{IK75}\\
$\Ot(n+(1/\eps)^3)$ & Kellerer and Pferschy~\cite{KP04}\\
$\Ot(n+(1/\eps)^{12/5})$ & Chan~\cite{Chan18}\\
$\Ot(n+(1/\eps)^{9/4})$ & Jin~\cite{Jin19}\\
$\Ot(n+(1/\eps)^{11/5})^\dagger$ & Deng, Jin, and Mao~\cite{DJM23}\\
$\Ot(n+(1/\eps)^2)$ & Chen, Lian, Mao, and Zhang~\cite{CLMZ25}; Mao~\cite{Mao24}\\
$\Ot(n+(1/\eps)^{1+\rho/2})^\dagger$ & This paper\\
\bottomrule
\end{tabular}
\end{table}

Recent integer 3SUM algorithms by Alman and Vassilevska
Williams~\cite{AV26} with reductions
to integer convolution, make it natural to revisit this barrier.
Using Wang's algorithm~\cite{Wang26}, Zhou~\cite{Zhou26} obtains witnessed integer
convolution with exponent $1.9601397040925$ and applies it to exact
Bounded Knapsack. We formulate our result as a transfer from an
integer convolution algorithm, so that the approximation argument is
independent of the implementation of that algorithm.

\begin{assumption}[Integer convolution]\label{ass:convolution}
Fix $1<\rho<2$. For arrays of length at most $m$ with entries in
$\mathbb Z\cup\{+\infty\}$, whose finite entries have at most $b$ bits,
their $(\min,+)$-convolution can be computed in
\[
 \Ot(m^\rho)\poly\bigl(b+\log(1/\eta)\bigr)
\]
bit operations, with failure probability at most $\eta$.
The guarantee holds for every input, using fresh randomness in each call.
\end{assumption}

\begin{theorem}\label{thm:main}
Under Assumption~\ref{ass:convolution}, Knapsack has a randomized FPTAS
running in
\[
 \Ot\left(n+\left(\frac1\eps\right)^{1+\rho/2}\right)
\]
time on inputs whose integers have magnitude polynomial in $n+1/\eps$.
For every fixed $a>0$, it returns a feasible solution within a factor
of $1+\eps$ of the optimum with probability at least
$1-(n+1/\eps+2)^{-a}$. 
\end{theorem}

For general binary inputs, the running time has
an additional factor polynomial in the maximum input bit length. Under the convolution guarantee in
\cite[Lemma~3.6]{Zhou26}, we have:

\begin{corollary}\label{cor:numerical}
Knapsack has a randomized FPTAS running in
$\Ot(n+(1/\eps)^{1.981})$ time, with the input and probability
guarantees of Theorem~\ref{thm:main}.
\end{corollary}
\begin{proof}
Substitute $\rho=1.9601397040925$ into Theorem~\ref{thm:main}.
Then $1+\rho/2=1.98006985204625<1.981$.
\end{proof}

\subsection{Technique overview}
We follow the functional approach of Chan~\cite{Chan18}. For an item
set $I$, let $f_I(x)$ be the maximum profit attainable at capacity $x$.
The reduction of Chen, Lian, Mao, and Zhang~\cite[Section~4]{CLMZ25} makes it
sufficient to approximate $f_I$ when the item profits are integers in
$[1/(2\eps),1/\eps]$ and
$f_I(t)\le 1/(\alpha\eps^2)$, for some $\alpha\in[1,1/\eps]$.
The permissible additive error is $\Ot(1/(\alpha\eps))$.

Label the items in decreasing order of efficiency, and let
$I_{\mathrm{head}}$ be the greedy prefix for capacity $t$.
The remaining items form $I_{\mathrm{tail}}$.
The proximity decomposition of Chen, Lian, Mao, and Zhang~\cite[Section~6]{CLMZ25}
has two useful properties. First, the tail can be divided into
logarithmically many groups, each of which contributes little profit
to an appropriate optimum at every capacity. Second, the capacity
range of the head can be divided into intervals, each with a common
high-, medium-, and low-efficiency part.

\paragraph{Requested prefixes and suffixes}
For the head, the high-efficiency parts are prefixes, and the
low-efficiency parts are suffixes. We approximate the profit omitted
from a high-efficiency part and the profit selected from a
low-efficiency part. Both quantities are bounded by $1/\eps^{3/2}$,
so relative accuracy $\Theta(\eps^{1/2}/\alpha)$ suffices.
Chan's blocked merging argument~\cite[Theorem~3]{Chan18}, with the
faster convolution primitive, combines $m$ uniform pseudo-concave
functions in $\Ot(m/\delta^{(\rho+1)/2})$ time at relative accuracy
$\delta$. We extend this computation to all requested prefixes and
suffixes by sharing a balanced tree of block functions.
The total cost for the head is $\Ot((1/\eps)^{1+\rho/2})$.

\paragraph{Exact computation on common profit grids}
For a tail group, normalize its profits to $[1,2]$, and let $B$ be
the relevant profit bound and $m$ the number of distinct profits.
Proximity gives $mB=\Ot(1/\eps)$. Chan's
rounding lemma~\cite[Lemma~6]{Chan18} assigns profits to
$\Ot(\sigma/\delta)$ grids of spacing between $\sigma$ and $2\sigma$,
with relative error $O(\delta)$. We compute each grid group exactly,
at total cost $\Ot(mB/\sigma)$, and then merge the grid groups in a
balanced tree, at cost $\Ot(\sigma/\delta^{\rho+1})$.
Balancing these two terms gives
$\Ot(\sqrt{mB}/\delta^{(\rho+1)/2})$.
Since the required inverse accuracy is $O(1/\eps)$, this also gives
$\Ot((1/\eps)^{1+\rho/2})$ time.

The reduction and proximity decomposition are used directly from
\cite{CLMZ25}. The changes to Chan's basic merging lemmas are given
in Appendix~\ref{app:merging}; the minor changes to the reduction
and the finite-precision grid construction are given in
Appendices~\ref{app:reduction} and~\ref{app:grids}, respectively.
The complement-and-merge argument for the head follows
\cite[Section~6.2]{CLMZ25} and is recorded in
Appendix~\ref{app:head}.
The main proof focuses on the common-grid computation and the
running-time bounds for the two parts.

\subsection{Further related work}
Chan, Itakura, Ma, and Yu~\cite[Corollary~2.4]{CIMY26} give a
simpler randomized algorithm for all-numbers integer 3SUM in
$O(m^{1.965})$ time. Combining its reporting guarantee with the
indexed-equality encoding and bit refinement
in~\cite[Appendix~B]{Zhou26} gives integer convolution exponent
$1.965$. This is larger than the exponent $1.9601397040925$
used in Corollary~\ref{cor:numerical}.

Proximity results have also led to faster exact algorithms for
Knapsack. Chen, Lian, Mao, and Zhang~\cite{CLMZ24} developed fine-grained proximity
bounds using additive combinatorics. Bringmann~\cite{Bri24} and
Jin~\cite{Jin24} subsequently obtained near-quadratic dependence on
the maximum item weight. These algorithms solve a specified
capacity; the functional approximation framework requires
approximations for all capacities. The simultaneous decomposition
of~\cite{CLMZ25} supplies this additional structure.

Bounded monotone convolution also yields algorithms for
Knapsack~\cite{BC22}. For length-$m$ monotone integer inputs with
values bounded by $O(m)$, Chi, Duan, Xie, and Zhang~\cite{CDXZ22}
give expected $\Ot(m^{3/2})$ time, and Jin, Park, Saha, and
Xu~\cite{JPSX26} give deterministic $m^{3/2+o(1)}$ time.
These smaller exponents require a numerical range bound.
Our inverse functions store exact original weights, whose range
is not controlled by the number of rounded profit values.
Consequently, these algorithms do not directly replace the general
integer convolution interface in Assumption~\ref{ass:convolution}.
For weak approximation, which permits weight at most $(1+\eps)t$,
Chen, Lian, Mao, and Zhang~\cite{CLMZ25weak} gave an
$\Ot(n+(1/\eps)^{7/4})$-time algorithm. The approximation scheme
studied here satisfies the original capacity constraint.

\subsection{Paper outline}
Section~\ref{sec:notation} fixes notation and the numerical model.
Section~\ref{sec:functional} gives the operations on functions.
Sections~\ref{sec:reduction} and~\ref{sec:proximity} recall the
reduction and proximity decomposition.
Section~\ref{sec:algorithm} presents the algorithm for the reduced
problem and proves Theorem~\ref{thm:main}.
We conclude in Section~\ref{sec:conclusion}.

\section{Notation}\label{sec:notation}
For an item set $J$, let $p(J)=\sum_{i\in J}p_i$ and
$w(J)=\sum_{i\in J}w_i$. The efficiency of item $i$ is
$p_i/w_i$. We write
\[
 f_J(x)=\max\{p(S):S\subseteq J,\ w(S)\le x\},\qquad
 h_J(x)=\min\{p(S):S\subseteq J,\ w(S)\ge x\}.
\]
Set $f_J(x)=-\infty$ for $x<0$, $h_J(x)=0$ for $x\le0$,
and $h_J(x)=+\infty$ for $x>w(J)$.
All algorithmic capacities and demands are integers. Packing functions
are extended to real capacities using their value at $\lfloor x\rfloor$.
Every covering approximation is zero for $x\le0$, with the empty witness.
A packing approximation needed only on $[0,t]$ is held constant
beyond $t$ when used in subsequent merges.

A monotone function $\hat f$ approximates $f$ from below with factor
$1+\delta$ if $\hat f\le f\le(1+\delta)\hat f$.
It has additive error $d$ if $\hat f\le f\le\hat f+d$.
For covering functions, approximation is from above:
$h\le\hat h\le(1+\delta)h$. Multiplicative statements concern
nonnegative finite values; zero values and the finite domain are
preserved. The complexity of a step function is its number of steps.
We use $\min(f,B)$ for pointwise capping at $B$.

Every reported pair has a witnessing subset. A packing pair $(x,y)$
has a witness of weight at most $x$ and profit at least $y$;
a covering pair has weight at least $x$ and profit at most $y$.
Reported values may therefore be rounded down for packing and up
for covering. Among pairs with the same reported value, keep the
smallest capacity coordinate for packing and the largest covered-demand
coordinate for covering. A covering pair certifies all demands at most
its coordinate; thus its witness remains valid throughout the resulting
constant-value interval.
Witnesses are stored by predecessor pointers, as in
\cite[Section~5]{Chan18}. Rounding and envelope operations retain the
existing pointer, and union nodes with an empty child are suppressed.
Only the final chosen witness is expanded.
When a function is translated vertically during a numerical
computation, the translation is undone before interpreting its
reported profits.

For arrays $A,B$, write
\[
 (A\star B)[k]=\min_{i+j=k}\{A[i]+B[j]\},
\]
where indices must belong to the input arrays. Witnesses need no
additional convolution assumption: replace $A[i]$ by $mA[i]+i$
and $B[j]$ by $mB[j]$, where both lengths are at most $m$.
Euclidean division of a finite answer by $m$ recovers its original
value and an attaining first index. This adds only $O(\log m)$ bits;
see also~\cite[Lemma~3.5]{Zhou26}. Negation gives the corresponding
$(\max,+)$ operation. We check each decoded index pair and recompute
its sum from the finite input entries. An out-of-range index or an
inconsistent value is discarded. This costs linear additional work
up to bit-length factors, and ensures that every retained entry has
a valid witness even if the convolution call errs. Approximation
guarantees are conditioned on the calls succeeding.

The notation $\Ot$ suppresses polylogarithmic factors in the input
size, the inverse accuracy, the ratio of positive function values,
and the inverse failure probability. In the applications these are
polylogarithmic in $n+1/\eps$, apart from the explicitly retained
bit-length factors. We may assume that $1/\eps$ is a power of four
and is at least four, by reducing $\eps$ within a constant factor. In particular,
$1/(2\eps)$ and $1/\eps^{3/2}$ are integers. All other accuracy and
grid parameters can be rounded to dyadic values within constant factors.

\section{Approximating for all capacities}\label{sec:functional}
For disjoint item sets $J,K$,
\[
 f_{J\cup K}=f_J\oplus f_K,\qquad
 (f\oplus g)(x)=\max_{0\le u\le x}\{f(u)+g(x-u)\}.
\]
The analogous identity for covering functions uses minimum instead
of maximum. A cover of demand $x$ can distribute that demand between
its two parts without assigning either part more than its weight.
We use this covering convolution only for nonnegative demands.

Approximation errors compose in the usual way
\cite[Section~3]{CLMZ25}: additive errors add, while a common
multiplicative factor is preserved by convolution.
Consequently, a balanced tree of approximate convolutions can use
accuracy $\delta/O(\log m)$ at each internal node to achieve total
factor $1+\delta$. Capping nonnegative packing functions also
preserves a common multiplicative factor.

The packing function of items having a common profit $p$ is
\emph{$p$-uniform and pseudo-concave}: its values are
$0,p,2p,\ldots$, and the differences between consecutive weight
breakpoints are nondecreasing. Indeed, these breakpoints are prefix
sums of the weights in increasing order. For covering, the maximum
weight attainable with $j$ such items is the sum of the $j$ largest
weights, a concave sequence. These are the structured inputs used below.

\subsection{Basic merging operations}
The functions below have integer weight coordinates and finite-bit
rational values. Whenever witnesses are combined, their underlying
item sets are disjoint. Empty item sets are handled by their exact
convolution identities.
The next lemma adapts Chan's basic approximation lemmas and
blocking argument to convolution exponent $\rho$
\cite[Lemmas~1--2 and Theorem~3]{Chan18}.
The arbitrary first function in part~\textup{(ii)} is already allowed
by the proof, as observed in~\cite[Lemma~3.2]{CLMZ25}.

\begin{lemma}\label{lem:merging}
Let $0<\delta\le1/2$. Under Assumption~\ref{ass:convolution},
the following operations retain witnesses and have output complexity
$\Ot(1/\delta)$.
\begin{enumerate}[label=\textup{(\roman*)},leftmargin=*]
\item Two nonnegative packing functions of total complexity $\ell$
can be convolved with approximation factor $1+\delta$ in
$\Ot(\ell+(1/\delta)^\rho)$ time.
If the second function is the packing function of a common-profit
item set, the time is $\Ot(\ell+1/\delta)$.
\item For $m\ge1$ packing functions of total complexity $\ell$, all
except possibly the first being common-profit item functions, their
convolution can be approximated with factor $1+\delta$ in
\[
 \Ot\left(\ell+\frac{m}{\delta^{(\rho+1)/2}}\right)
\]
time.
\end{enumerate}
The same bounds hold for covering functions, with approximation
from above and common-profit covering functions as the structured inputs.
\end{lemma}

The changes to~\cite{Chan18} are the substitution of the integer
convolution algorithm and the covering version. We give the details
in Appendix~\ref{app:merging}, including exact weight coordinates
and the treatment of a truncated last uniform step.

We will also use the following elementary consequence. If the finite
values of two packing functions lie in intervals of lengths $B_1$
and $B_2$, subtract the two lower endpoints, apply
Lemma~\ref{lem:merging}\textup{(i)}, and restore their sum.
The additive error is $O(\delta(B_1+B_2))$.
Translations affect only finite values, so this operation also
applies to functions that are undefined at small capacities.

\subsection{Requested prefixes and suffixes}
We need a version of the blocking argument that shares computation
between several nested item sets.

\begin{lemma}\label{lem:prefixes}
For a sequence of $N$ items and $q$ requested prefixes, approximations
of all the requested packing functions with factor $1+\delta$ can
be computed in
\[
 \Ot\left(
   \frac{N}{\delta^{(\rho+1)/2}}
   +\frac{q+1}{\delta^\rho}
 \right)
\]
time. Each output has complexity $\Ot(1/\delta)$ and retains
witnesses. The statement also holds for requested suffixes and for
covering functions.
\end{lemma}
Use Chan's blocks, cut additionally at requested boundaries, and store
their convolutions in a balanced binary tree. Each requested prefix
is a disjoint union of logarithmically many tree nodes.
This extension of the blocking argument is proved in
Appendix~\ref{app:prefixes}.

\subsection{Exact computation within common-grid groups}
The next lemma uses the grid family of~\cite[Lemma~6]{Chan18}.
The change is to compute each grid group independently and exactly,
then approximately merge the resulting functions.

\begin{lemma}\label{lem:grids}
Let $J$ contain $N$ items with positive integer weights and at most
$m$ distinct rational profits in $[1,2]$. Let $B\ge2$ and
$0<\delta\le1/2$ be rational.
Under Assumption~\ref{ass:convolution}, a witnessed approximation
of $\min(f_J,B)$ with factor $1+\delta$ and complexity
$\Ot(1/\delta)$ can be computed with high probability in time
\begin{equation}\label{eq:grid-bound}
 \Ot\left(
 N+mB+\frac1{\delta^\rho}
       +\frac{\sqrt{mB}}{\delta^{(\rho+1)/2}}
 \right).
\end{equation}
The grids have finite-bit rational spacings, and the convolution
entries are exact integer weights.
\end{lemma}
\begin{proof}
We may reduce $\delta$ by an absolute constant throughout the
construction; it suffices to obtain factor $1+O(\delta)$.

\paragraph{Truncating equal-profit classes}
For a profit $p$, keep only the
$\min\{|J_p|,\lfloor(B+2)/p\rfloor\}$ lightest items in its class.
This preserves $\min(f_J,B)$. A solution of profit below $B$ uses
at most this many copies of every profit. A solution reaching $B$
has a subset of profit in $[B,B+2]$, obtained by stopping as soon as
the accumulated profit reaches $B$. Replace each class selection
by equally many lightest items. Thus only $O(mB)$ items need to be
retained, at preprocessing cost $\Ot(N+mB)$.

\paragraph{Assigning profit grids}
For a parameter $\sigma$ with $\delta\le\sigma\le1/4$,
Chan's lemma constructs, in $\Ot(1/\delta)$ time, a family
\[
 \mathcal D\subseteq[\sigma,2\sigma],
 \qquad |\mathcal D|=\Ot(\sigma/\delta),
\]
such that every profit $p\in[1,2]$ can be rounded down to a
multiple $p'=a_pd$ of some $d\in\mathcal D$, with
$p-p'=O(\delta)$. Assign an entire profit class to one such $d$.
This loses only a factor $1+O(\delta)$, since $p\ge1$.
Appendix~\ref{app:grids} gives the finite-bit implementation and
the verification of the grid family.

\paragraph{Computing one grid group exactly}
Let $J_d$ be the retained items assigned to $d$, and let $m_d$ be
their number of profit classes. Put $T=\lceil(B+2)/d\rceil$.
Compute the exact minimum-weight table for rounded profit $kd$,
for $0\le k\le T$, starting with $D(0)=0$ and $D(k)=+\infty$
for $k>0$.
If a class has rounded profit $ad$ and $W(i)$ is the sum of its
$i$ lightest weights, adding the class updates the table by
\begin{equation}\label{eq:grid-update}
 D_{\mathrm{new}}(k)
    =\min_i\{D_{\mathrm{old}}(k-ai)+W(i)\}.
\end{equation}
Here $i$ ranges over all available integer counts with $ai\le T$,
and a negative old-table index has value $+\infty$.
All rounded profits are positive, so discarding states above $T$
cannot affect the entries being computed.
For each residue modulo $a$, this is convolution with a convex
sequence. The linear-time operation of~\cite[Lemma~9]{AT19},
equivalently~\cite[Fact~1(ii$'$)]{Chan18}, computes all residues
in $O(T)$ time, retaining attaining counts: each residue uses
$O(T/a+1)$ entries, and $a\le2/d\le T$.
The full count sequence needed through $T$ is retained;
no fractional last step is inserted.
All updates for this grid cost $\Ot(m_dB/d)$.

The table determines $\min(f'_{J_d},B)$ exactly, where primes denote
rounded profits. The first subset reaching $B$ has rounded profit
at most $B+2$, and therefore appears in the table.
Sorting the weight-profit pairs and taking their upper envelope
costs $\Ot(B/d)$. Summing over nonempty grid groups gives
\[
 \Ot\left(\sum_{d\in\mathcal D}\frac{m_dB}{d}\right)
   =\Ot(mB/\sigma).
\]

\paragraph{Merging the grid groups}
Compress each function with relative error $O(\delta)$ and merge
the functions in a balanced tree using
Lemma~\ref{lem:merging}\textup{(i)}, with the local accuracy divided
by the tree depth. Cap intermediate packing values at $B$.
This is valid because, for nonnegative $v_i$,
\[
 \min\left\{\sum_i\min\{v_i,B\},B\right\}
   =\min\left\{\sum_i v_i,B\right\}.
\]
The merges cost
$\Ot(|\mathcal D|/\delta^\rho)=\Ot(\sigma/\delta^{\rho+1})$,
and their total factor is $1+O(\delta)$.
Every reported profit is certified by original items, since profits
were rounded down and the groups are disjoint.

Choose $\sigma$ within a constant factor of
$\sqrt{mB\,\delta^{\rho+1}}$ when this lies in $[\delta,1/4]$.
Both costs then equal
$\Ot(\sqrt{mB}/\delta^{(\rho+1)/2})$.
If this value is below $\delta$, round to a single grid of spacing
$\Theta(\delta)$ and compute its table exactly. Its cost is
$\Ot(mB/\delta)=\Ot(1/\delta^\rho)$ in this case.
If it is above $1/4$, take $\sigma=1/4$; the cost is $\Ot(mB)$.
These cases, together with preprocessing, prove~\eqref{eq:grid-bound}.
\end{proof}

\section{Reducing the problem}\label{sec:reduction}
We use the reduced problem $\mathrm{RP}(\eps,\alpha)$ of
\cite[Section~4]{CLMZ25}:
\begin{quote}
Given $n$ items and a capacity $t$, assume that their profits are
integers in $[1/(2\eps),1/\eps]$ and that
$f_I(t)\le1/(\alpha\eps^2)$ for some $\alpha\in[1,1/\eps]$.
Compute a witnessed monotone approximation of $f_I$ on $[0,t]$
with additive error $\Ot(1/(\alpha\eps))$ and complexity
$\Ot(1/\eps)$.
\end{quote}
An error with the displayed polylogarithmic factors is sufficient:
reducing the internal precision parameter by a polylogarithmic
factor gives the desired final accuracy.
It suffices to handle powers of two $\alpha$. Replacing $\alpha$
by the largest such value not exceeding it weakens the optimum
bound and changes the required additive error by at most a factor
of two.

\begin{lemma}[Adapted from {\cite[Section~4]{CLMZ25}}]\label{lem:reduction}
Suppose $\mathrm{RP}(\eps,\alpha)$ can be solved in
$\Ot(n+(1/\eps)^\beta)$ time, uniformly for
$1\le\alpha\le1/\eps$, where $\beta\ge\rho$ is fixed.
Under Assumption~\ref{ass:convolution}, Knapsack then has an
$\Ot(n+(1/\eps)^\beta)$-time FPTAS with the same high-probability
guarantee.
\end{lemma}

The reduction scales profits, bundles tiny items, and partitions the
remaining items into logarithmically many profit bands. The only
running-time change to~\cite{CLMZ25} is that the band functions are
merged in $\Ot((1/\eps)^\rho)$ time using
Lemma~\ref{lem:merging}\textup{(i)}.
For completeness, the adapted proof is in Appendix~\ref{app:reduction}.

\section{A proximity decomposition}\label{sec:proximity}
Fix an instance of $\mathrm{RP}(\eps,\alpha)$ and label its items in
decreasing order of efficiency. Let $b$ be the first index with
$\sum_{i=1}^b w_i>t$, or $n+1$ if all items fit.
Set
\[
 I_{\mathrm{head}}=\{1,\ldots,b-1\},\qquad
 I_{\mathrm{tail}}=\{b,\ldots,n\}.
\]
Since the head is feasible and every profit is at least $1/(2\eps)$,
\begin{equation}\label{eq:head-size}
 |I_{\mathrm{head}}|\le\frac{2}{\alpha\eps},
 \qquad p(I_{\mathrm{head}})\le\frac1{\alpha\eps^2}.
\end{equation}

We quote the following consequences of the proximity result
in~\cite[Lemma~5.1 and Section~6]{CLMZ25}. This is the structural
part of that algorithm; we use its stated simultaneous guarantees.

\begin{lemma}[Proximity decomposition {\cite{CLMZ25}}]\label{lem:proximity}
Let $\tau=\lceil c(1/\eps)^{1/2}\log(2/\eps)\rceil$, for a
sufficiently large absolute constant $c$.
The following decompositions can be constructed in $\Ot(n)$ time.
\begin{enumerate}[label=\textup{(\roman*)},leftmargin=*]
\item The tail is partitioned into $I_0,\ldots,I_r$, where
$r=O(\log(2/\eps))$ and $I_j$ has at most $2^j\tau$ distinct
profits. For every $x\le t$, some optimal solution $S$ for $(I,x)$
satisfies, simultaneously for all $j\ge1$,
\[
 p(S\cap I_j)\le\frac1{2^{j-1}\eps^{3/2}}.
\]
Every nonempty group with $j\ge1$ has $2^{j-1}\tau\le1/\eps$.
In particular, on $[0,t]$,
\begin{equation}\label{eq:tail-decomposition}
 f_I=f_{I_{\mathrm{head}}}\oplus f_{I_0}
       \oplus\bigoplus_{j=1}^r
       \min\left(f_{I_j},\frac1{2^{j-1}\eps^{3/2}}\right).
\end{equation}
\item If $\alpha\le1/\eps^{1/2}$, there are
$q=O(1/(\alpha\eps^{1/2}))$ capacity intervals covering $[0,t]$.
For each interval $j$, the head is partitioned into a prefix
$I_H^j$, a middle part $I_M^j$, and a suffix $I_L^j$.
The middle part has at most $3\tau$ distinct profits, and
\begin{equation}\label{eq:middle-size}
 \sum_{j=1}^q|I_M^j|\le\frac6{\alpha\eps}.
\end{equation}
For every capacity in interval $j$, there is an optimal solution
$S$ for the head at that capacity such that
\begin{equation}\label{eq:head-proximity}
 p(I_H^j\setminus S)\le\frac1{\eps^{3/2}},
 \qquad p(S\cap I_L^j)\le\frac1{\eps^{3/2}}.
\end{equation}
\end{enumerate}
\end{lemma}

Part~\textup{(i)} is the tail partition and
Corollary~6.2 of~\cite{CLMZ25}. For a nonempty $I_j$, the preceding
tail prefix has $2^{j-1}\tau$ distinct profits, giving the additional
displayed bound. Part~\textup{(ii)} is the interval partition of
\cite[Section~6.2, Lemmas~6.6--6.10]{CLMZ25}; each middle part
consists of at most three consecutive blocks. The additive constant
in the number of intervals is absorbed because
$\alpha\le1/\eps^{1/2}$.

For clarity, the capped identity~\eqref{eq:tail-decomposition} follows
directly from the simultaneous guarantee. Its right-hand side is
at most $f_I$, since capping only lowers packing values.
For the reverse inequality, give each part the weight of its
intersection with the stated optimal solution. Every cap is then
at least that part's contribution.

\section{The algorithm}\label{sec:algorithm}
We now solve $\mathrm{RP}(\eps,\alpha)$ using
Lemma~\ref{lem:proximity}. We first handle the tail and then the head,
following the organization of~\cite[Section~6]{CLMZ25}.

\subsection{Tackling the tail}
\begin{lemma}\label{lem:tail}
The functions $f_{I_0}$ and
$\min(f_{I_j},1/(2^{j-1}\eps^{3/2}))$ for $j\ge1$ have witnessed
approximations of complexity $\Ot(1/\eps)$ each and additive error
$O(1/(\alpha\eps))$ each. All these approximations can be computed
with high probability in
$\Ot(n+(1/\eps)^{1+\rho/2})$ total time.
\end{lemma}
\begin{proof}
For $I_0$, group items by their at most $\tau$ distinct profits.
Apply Lemma~\ref{lem:merging}\textup{(ii)} with
$\delta=\Theta(\eps)$ to the common-profit functions.
Since $f_{I_0}(t)\le1/(\alpha\eps^2)$, the additive error is
$O(1/(\alpha\eps))$. The cost is
\[
 \Ot\left(|I_0|+\frac{\tau}{\eps^{(\rho+1)/2}}\right)
   =\Ot\left(|I_0|+\left(\frac1\eps\right)^{1+\rho/2}\right).
\]

Consider a nonempty $I_j$ with $j\ge1$. Divide its profits by
$1/(2\eps)$, so that they lie in $[1,2]$, and use the cap
\begin{equation}\label{eq:tail-cap}
 B=\min\left\{
       \frac{2}{2^{j-1}\eps^{1/2}},
       \frac{2}{\alpha\eps}
       \right\}.
\end{equation}
The second bound is valid because $f_{I_j}(x)\le f_I(t)$ for $x\le t$.
Moreover, $B\ge2$: for its first term, use
$2^{j-1}\tau\le1/\eps$ and $\tau\ge1/\eps^{1/2}$;
for the second, use $\alpha\le1/\eps$.
If $m$ is the number of distinct profits in $I_j$, then
\begin{equation}\label{eq:tail-product}
 mB\le 2^j\tau\cdot\frac{2}{2^{j-1}\eps^{1/2}}
      =\frac{4\tau}{\eps^{1/2}}=\Ot(1/\eps).
\end{equation}
Apply Lemma~\ref{lem:grids} with
$\delta=\Theta(1/(\alpha B))$.
Its additive error is $O(\delta B)=O(1/\alpha)$ in normalized
profits, hence $O(1/(\alpha\eps))$ in the original units.
Also, $1/\delta=O(\alpha B)=O(1/\eps)$.
By~\eqref{eq:grid-bound} and~\eqref{eq:tail-product}, the time is
\[
 \Ot\left(
 |I_j|+\frac1\eps+\left(\frac1\eps\right)^\rho
       +\frac1{\eps^{1/2}}\left(\frac1\eps\right)^{(\rho+1)/2}
 \right)
 =\Ot\left(|I_j|+\left(\frac1\eps\right)^{1+\rho/2}\right).
\]
There are logarithmically many disjoint tail groups, so these costs
give the asserted total.
\end{proof}

\subsection{Tackling the head}
\begin{lemma}\label{lem:head}
The function $f_{I_{\mathrm{head}}}$ has a witnessed approximation
with additive error $O(1/(\alpha\eps))$ and complexity
$\Ot(1/\eps)$, computable with high probability in
$\Ot((1/\eps)^{1+\rho/2})$ time.
\end{lemma}
\begin{proof}
If the head is empty, its function is zero.
If $\alpha\ge1/\eps^{1/2}$, apply
Lemma~\ref{lem:merging}\textup{(ii)} to its singleton functions
with accuracy $\Theta(\eps)$.
By~\eqref{eq:head-size}, the error is $O(1/(\alpha\eps))$
and the cost is
\[
 \Ot\left(\frac1{\alpha\eps}\cdot
                 \frac1{\eps^{(\rho+1)/2}}\right)
 \le \Ot\left(\left(\frac1\eps\right)^{1+\rho/2}\right).
\]
Henceforth assume $\alpha<1/\eps^{1/2}$ and use the $q$ intervals
of Lemma~\ref{lem:proximity}\textup{(ii)}.

\paragraph{High- and low-efficiency parts}
Set $\delta=\Theta(\eps^{1/2}/\alpha)$, with a sufficiently small
constant. Lemma~\ref{lem:prefixes} computes upper approximations
$\hat h_{I_H^j}$ for all requested prefixes and lower approximations
$\hat f_{I_L^j}$ for all requested suffixes, each with factor
$1+\delta$.
By~\eqref{eq:head-size}, the total cost is
\[
 \Ot\left(
 \frac1{\alpha\eps}
       \left(\frac{\alpha}{\eps^{1/2}}\right)^{(\rho+1)/2}
 +(q+1)\left(\frac{\alpha}{\eps^{1/2}}\right)^\rho
 \right)
 =\Ot\left(\left(\frac1\eps\right)^{1+\rho/2}\right).
\]
Indeed, the first term is largest at $\alpha=1/\eps^{1/2}$
and equals $(1/\eps)^{1+\rho/2}$ there; the second is
$O((1/\eps)^\rho)$ at that endpoint.
The total output complexity is
$\Ot(q/\delta)=\Ot(1/\eps)$.

Use the complement-and-merge construction of
\cite[Lemmas~6.12--6.13]{CLMZ25}: cap the suffix profit at
$1/\eps^{3/2}$, retain prefix covering losses through
$(1+\delta)/\eps^{3/2}$, complement the prefix witnesses, and
merge the two resulting functions after translating their finite
value ranges. Appendix~\ref{app:head} verifies this adaptation,
including the enlarged covering cutoff.
For each interval it gives a witnessed function $g_j$ with error
$O(1/(\alpha\eps))$ on the contributions permitted by proximity.
Its finite values lie in
\[
 \left[p(I_H^j)-O(1/\eps^{3/2}),
       \ p(I_H^j)+1/\eps^{3/2}\right].
\]
The $q$ merges cost $\Ot(q/\delta^\rho)$ and their total
complexity is $\Ot(q/\delta)=\Ot(1/\eps)$.

\paragraph{Medium-efficiency parts}
Translate the finite values of $g_j$ into
$[0,O(1/\eps^{3/2})]$. Group $I_M^j$ by profit into at most
$3\tau$ common-profit functions and merge them with the translated
$g_j$ using Lemma~\ref{lem:merging}\textup{(ii)} at accuracy
\[
 \Theta\left(
   \frac1{\alpha(|I_M^j|+1/\eps^{1/2})}
 \right).
\]
The maximum translated value is
$O(1/\eps^{3/2}+|I_M^j|/\eps)$, so the added error is
$O(1/(\alpha\eps))$. Restore the translation and call the result
$q_j$. By~\eqref{eq:middle-size},
\begin{equation}\label{eq:head-precision-sum}
 \sum_{j=1}^q\alpha\bigl(|I_M^j|+1/\eps^{1/2}\bigr)
 \le O(1/\eps).
\end{equation}
Thus the total output complexity is $\Ot(1/\eps)$.
Since $(\rho+1)/2>1$, the sum of the $(\rho+1)/2$ powers of
these inverse accuracies is at most the $(\rho+1)/2$ power of
their sum. The total middle-part cost is therefore
\[
 \Ot\left(
 \frac1{\eps^{1/2}}\sum_{j=1}^q
   \bigl[\alpha(|I_M^j|+1/\eps^{1/2})\bigr]^{(\rho+1)/2}
 \right)
 =\Ot\left(\left(\frac1\eps\right)^{1+\rho/2}\right).
\]
Reading the functions and the middle parts is covered by the same
bound. The term $1/\eps^{1/2}$ in the accuracy also covers short
or empty middle parts.

Return $\max\{0,q_1,\ldots,q_q\}$ with the selected witnesses.
By the proximity guarantee and the assembly argument of
\cite[Lemmas~6.14--6.15]{CLMZ25}, detailed in
Appendix~\ref{app:head}, this approximates
$f_{I_{\mathrm{head}}}$ with additive error $O(1/(\alpha\eps))$.
The upper envelope takes time near-linear in the total complexity.
\end{proof}

\subsection{Putting things together}
\begin{lemma}\label{lem:reduced-solver}
Under Assumption~\ref{ass:convolution},
$\mathrm{RP}(\eps,\alpha)$ can be solved with high probability in
$\Ot(n+(1/\eps)^{1+\rho/2})$ time, uniformly for
$1\le\alpha\le1/\eps$.
\end{lemma}
\begin{proof}
Construct the decomposition in Lemma~\ref{lem:proximity}.
Use Lemmas~\ref{lem:tail} and~\ref{lem:head} to approximate its
component functions. By~\eqref{eq:tail-decomposition}, their exact
convolution equals $f_I$ on $[0,t]$.
The component approximation errors add to
$\Ot(1/(\alpha\eps))$.

Use a balanced tree of Lemma~\ref{lem:merging}\textup{(i)} to merge
the $O(\log(2/\eps))$ functions with total factor $1+O(\eps)$.
The work is $\Ot((1/\eps)^\rho)$.
At capacities at most $t$, all reported packing values are at most
$f_I(t)\le1/(\alpha\eps^2)$, so the additional additive error is
$O(1/(\alpha\eps))$. The output complexity is $\Ot(1/\eps)$.
Since $\rho<1+\rho/2$, this final work is within the claimed bound.
\end{proof}

\begin{proof}[Proof of Theorem~\ref{thm:main}]
Apply Lemma~\ref{lem:reduction} with
$\beta=1+\rho/2\ge\rho$ and the solver of
Lemma~\ref{lem:reduced-solver}.
The reduced errors, the logarithmically many profit bands, and the
merge depths contribute only fixed polylogarithmic factors.
Run the reduction at precision
$\eps/(C\log^c(n+1/\eps+2))$, rounded down to a reciprocal power
of four, for suitable absolute constants $C,c$.
The total loss is then at most $\eps/(1+\eps)$ times the optimum.
This changes the running time only by polylogarithmic factors.

For a prescribed total failure probability
$\zeta=(n+1/\eps+2)^{-a}$, assign the $s$-th randomized call budget
$\zeta/(4(s+1)^2)$. Each call uses fresh randomness, so conditioning
on its input history gives the required bound even for adaptive
calls. Summing these budgets bounds the probability of any failure
by $\zeta$. Grid coverage is verified before use, as in
Appendix~\ref{app:grids}.

On the event that all calls succeed, witness pointers give a
subset whose original weight is at most $t$ and whose original
profit is at least the reported profit. The selected head
representation uses one interval only, and all unions within that
representation are over disjoint item sets. The pointer convention
in Section~\ref{sec:notation} leaves only binary unions of nonempty
witnesses and the selected prefix complements. Within each profit
band there is at most one such complement, which is expanded by a
scan of its prefix. Expanding the witness, including the tiny-item
bundles, therefore takes $O(n)$ time.
The final subset is checked against the original input; a detected
failure returns the empty set, so feasibility is always preserved.
\end{proof}

\paragraph{Bit complexity}
Let $\mathcal L$ be the maximum bit length of an input integer plus
$\lceil\log_2(n+1/\eps+2)\rceil$.
Inverse-array coordinates are exact integers derived from the original
weights. Their sums, negations, witness perturbations, and the finite
Monge penalties in Appendix~\ref{app:merging} all have $O(\mathcal L)$ bits.
We implement compression by dyadic value bands with dyadic additive
meshes, rather than storing long exact powers of $1+\delta$.
The grids of Appendix~\ref{app:grids} also use dyadic rationals.
Profit scaling and grid alignment therefore use numbers of encoding
length polynomial in $\mathcal L$.
The overall bit complexity is
\[
 \Ot\left(n+\left(\frac1\eps\right)^{1+\rho/2}\right)
 \poly(\mathcal L).
\]
For polynomially bounded input integers, the additional factor is
polylogarithmic and is absorbed into $\Ot$.

\section{Conclusion}\label{sec:conclusion}
We have shown that integer convolution exponent $\rho<2$ gives a
Knapsack FPTAS with exponent $1+\rho/2$ in $1/\eps$.
The saving is $(2-\rho)/2$.
The all-capacities decomposition of~\cite{CLMZ25} reduces the problem
to nested head functions and capped tail functions. Shared block
computations handle the former; exact common-grid computations
followed by approximate merging handle the latter.
Both fit within the same bound. Improving the transfer exponent
requires a better bound for these computations, or a different
decomposition.

\section*{AI Disclosure}
This result was developed by GPT with little human input and the authors take no intelectual credit.

\appendix
\section{Adaptations of Chan's merging lemmas}\label{app:merging}
This appendix proves Lemma~\ref{lem:merging}. The packing construction
follows~\cite[Lemmas~1--2 and Theorem~3]{Chan18}.
We include the adaptations for integer convolution, upward covering
approximation, and restricted finite domains.

\subsection{Two-function merging}
Fix a positive dyadic value scale $v$ and a mesh
$\Delta=\Theta(\delta v)$. All arrays below have $O(1/\delta)$
entries; a constant extension of the index range accommodates
rounding and the sum of two inputs.

\paragraph{Packing}
Cap the two input functions at $v$ and round values down to
multiples of $\Delta$. For an input $F$, form the threshold inverse
\[
 A_F[k]=\min\{x:F(x)\ge k\Delta\},
\]
using $+\infty$ for an unattainable threshold.
For $k=0$, this is the first point of the finite domain, which need
not be zero for a translated, restricted function.
Keep the witness of the corresponding breakpoint.
The $(\min,+)$-convolution of the inverse arrays gives the least
weight attaining each combined rounded profit threshold.
Each finite answer is certified by the union of its two witnesses.
The rounding loses at most $2\Delta$.
After capping at $v$, this approximates the capped convolution
with additive error $O(\delta v)$.
The array entries are exact integer weights, so
Assumption~\ref{ass:convolution} gives time
$\Ot(\ell+(1/\delta)^\rho)$.

\paragraph{Covering}
For a covering function $F$, use its budget inverse
\[
 A_F[k]=\max\{x:F(x)\le k\Delta\}.
\]
Here $x$ is an integer demand within the stored finite weight
range, and $A_F[0]$ includes demand zero.
The $(\max,+)$-convolution of these arrays gives the greatest
covered demand at each combined profit budget.
If its output is $C[k]$, report
\[
 \hat h(x)=\min\{k\Delta:C[k]\ge x\},
\]
with value $+\infty$ when the set is empty. Use the witness
attaining $C[k]$, which covers the entire interval of demands
assigned to that budget. In particular, compression of equal
reported budgets keeps the greatest covered-demand coordinate.
If an exact combined covering loss is at most $v$, rounding its
two component budgets up raises their sum by at most $2\Delta$.
No weight or demand is rounded. Negating the arrays permits use
of Assumption~\ref{ass:convolution}.

\paragraph{A common-profit second input}
Suppose the second input consists of items of profit $p$.
Let $W(j)$ be the sum of their $j$ smallest weights and $V(j)$
the sum of their $j$ largest weights.
Then $W$ is convex and $V$ is concave. Convolution with these
sequences is linear-time by Monge matrix searching
\cite{AKMSW87,AT19}; residue-class decomposition gives the
common-grid operation in~\cite[Fact~1(ii$'$)]{Chan18}.

Missing entries and finite count intervals can be handled by a finite
Monge completion. After negating in the concave case, let $M\ge1$
bound the absolute finite entries of both sequences.
Replace missing entries of the arbitrary sequence by $8M$ and extend
the convex sequence outside its count interval with slopes $-8M$
on the left and $8M$ on the right. Convexity is preserved.
Every valid candidate sum is in $[-2M,2M]$, whereas every invalid
candidate exceeds $2M$. Thus the ordinary Monge search applies,
and an answer above $2M$ is marked unattainable.
The extension adds only logarithmically many bits in the index range.

If $p>v$, any packing choice of a structured item already reaches
the cap $v$. It suffices to compare the first input alone with the
lightest structured item together with the first finite witness
of the first input, reporting value $v$ for the latter.
The first input's numerical contribution may be lowered to zero
in this candidate. For covering losses at most $v$, no structured
item is needed.

If $\delta v\le p\le v$, put
$a=\lceil p/(\delta v)\rceil$ and $\Delta=p/a$.
The structured input occupies indices $aj$.
Split the other array into residues modulo $a$ and convolve each
with $W$ for packing or $V$ for covering.
There are $O(1/\delta)$ entries over all residue arrays and
structured copies. In packing, keep complete counts through
the first count whose profit exceeds $v$, and cap only after
convolution. This avoids changing the final convex increment.
Use only available counts, retaining all of them if their total
profit is below $v$.

If $p<\delta v$, put $a=\lceil\delta v/p\rceil$ and
$\Delta=ap$. For packing, replace a count $j$ by
$a\lfloor j/a\rfloor$, losing less than $\Delta$, and use the
convex sequence $W(ak)$.
For covering, budget $k\Delta$ uses
$\min\{ak,s\}$ items, where $s$ is the class size, and covers
weight $V(\min\{ak,s\})$.
This sequence is concave: full blocks have nonincreasing weights,
the last partial block is no heavier than its predecessor, and
later increments are zero.
Rounding a required count up to the next block, capped at $s$,
increases the budget by less than $\Delta$ and does not decrease
the covered demand. These operations take
$\Ot(\ell+1/\delta)$ time and have additive error $O(\delta v)$.
Interpret the resulting entries as weight-value pairs and form
their monotone envelope. In particular, sparse count-index tables
need not themselves be monotone threshold inverses.

\paragraph{Combining scales}
Apply the procedure at all dyadic scales in the positive value
range, including a scale at least the largest possible output.
Take the upper envelope for packing and the lower envelope for
covering. A scale with the exact value in $(v/2,v]$ gives relative
error $O(\delta)$.
Handle zero values and finite-domain endpoints exactly.
Compression can use dyadic value bands, each with an additive
mesh proportional to $\delta$ times the band's lower endpoint.
It preserves the direction of approximation and produces
$\Ot(1/\delta)$ steps with short rational descriptions.
Changing the internal accuracy by a constant factor gives
the factor $1+\delta$ in
Lemma~\ref{lem:merging}\textup{(i)}.

\subsection{The blocked computation}
This is the proof of~\cite[Theorem~3]{Chan18} with the general
merge cost replaced by $(1/\delta)^\rho$.
Let $k=\lceil(1/\delta)^{(\rho-1)/2}\rceil$ and divide the
$m$ input functions into consecutive blocks of at most $k$ functions.
Only the first block can contain an arbitrary first function.

In a block of $s$ functions, merge incrementally at local accuracy
$\Theta(\delta/s)$. Every second input is structured.
The accumulated relative error is $O(\delta)$, and the work is
$\Ot(\ell_{\mathrm{block}}+s^2/\delta)$.
Compress the result to complexity $\Ot(1/\delta)$.
If $m\ge k$, merge the $O(m/k)$ block functions in a balanced
tree, at accuracy $\Theta(\delta/\log(m+2))$ per level.
The two costs are
\[
 \Ot\left(\ell+\frac{mk}{\delta}\right)
 \quad\text{and}\quad
 \Ot\left(\frac{m/k}{\delta^\rho}\right),
\]
both bounded by
$\Ot(\ell+m/\delta^{(\rho+1)/2})$.
If $m<k$, a single block costs
$\Ot(\ell+m^2/\delta)\le
\Ot(\ell+mk/\delta)$, and no final tree is needed.
Dividing the error budgets by suitable constants gives factor
$1+\delta$ overall. The same argument works from above for
covering. This proves Lemma~\ref{lem:merging}\textup{(ii)}.

\subsection{Sharing blocks between requested prefixes}\label{app:prefixes}
We prove Lemma~\ref{lem:prefixes} by extending the preceding
blocking construction.
Let $k=\lceil(1/\delta)^{(\rho-1)/2}\rceil$.
Cut the sequence into blocks of at most $k$ items, and cut additionally
at every requested boundary. There are $O(N/k+q+1)$ blocks.
For a block of $s\le k$ items, the incremental procedure of
\cite[Lemma~2(ii)]{Chan18}, or the structured operation in
Lemma~\ref{lem:merging}\textup{(i)} at accuracy $\Theta(\delta/s)$,
computes its function in $\Ot(s^2/\delta)$ time.
Compress each block function to complexity $\Ot(1/\delta)$.
The total block cost is $\Ot(Nk/\delta)$.

Build a balanced binary tree of the blocks, and compute each internal
node by Lemma~\ref{lem:merging}\textup{(i)}.
Every requested prefix is the disjoint union of
$O(\log(N+q+2))$ tree nodes. Merge their functions to answer the
request. There are $\Ot(N/k+q+1)$ merges in total, and every
computation path has logarithmic depth. Reducing local accuracies
by this depth, and assigning a constant fraction of the total
error to the block computations, gives factor $1+\delta$.
The cost is
\[
 \Ot\left(\frac{Nk}{\delta}
       +\frac{N/k+q+1}{\delta^\rho}\right)
 =
 \Ot\left(\frac{N}{\delta^{(\rho+1)/2}}
       +\frac{q+1}{\delta^\rho}\right).
\]
Reversing the sequence handles suffixes.
The covering version uses the covering operations throughout.
The empty item sequence is handled by its exact function.

\section{The modified profit-band reduction}\label{app:reduction}
We prove Lemma~\ref{lem:reduction}.
The argument is the reduction of~\cite[Section~4]{CLMZ25};
the faster general merge is the only change in its running-time
analysis.

Discard items of zero profit and items heavier than $t$.
If no items remain, return the empty set.
The standard greedy-prefix and best-single-item algorithm gives a
feasible profit $y$ with $y\le f_I(t)\le2y$
\cite[Section~2.5]{KPP04}.
Divide all profits by $y\eps^2$.
The scaled optimum lies in $[1/\eps^2,2/\eps^2]$, and it suffices
to incur additive error $\Ot(1/\eps)$.

\paragraph{Tiny items}
Order all items with scaled profit at most $1/\eps$ by decreasing
efficiency. Bundle consecutive items until each completed bundle
has profit in $(1/\eps,2/\eps]$, and discard a final incomplete
bundle of profit at most $1/\eps$. Replace a completed bundle by
one item of the same total profit and weight.
This costs at most $3/\eps$ profit at every capacity.
Indeed, for a feasible tiny-item set $S$, consider the longest
prefix of completed bundles of weight at most $w(S)$.
The fractional greedy solution at weight $w(S)$ dominates $S$.
The completed prefix misses at most the crossing bundle, of
profit at most $2/\eps$, and the discarded final bundle.
Each bundle has a witness consisting of its original items.

\paragraph{Profit bands}
The remaining profits belong to $(1/\eps,2/\eps^2]$.
Form bands $(\alpha/\eps,2\alpha/\eps]$, where $\alpha$ ranges
over powers of two up to $1/\eps$.
This gives $O(\log(2/\eps))$ bands.
Within a band, divide profits by $2\alpha$ and round them down
to integers. The profits now lie in $[1/(2\eps),1/\eps]$,
and the optimum is at most $1/(\alpha\eps^2)$.
Thus the band is an instance of $\mathrm{RP}(\eps,\alpha)$.

Every feasible solution uses at most $2/(\alpha\eps)$ items from
the band, so rounding loses at most that many normalized units.
A reduced-function error of $\Ot(1/(\alpha\eps))$ therefore
becomes $\Ot(1/\eps)$ after restoring the band's scale.

\paragraph{Combining the bands}
Compute each reduced function and restore its scale.
Replace any negative reported packing value by zero with the
empty witness before merging the bands.
The bands partition the items, so their input costs sum to
$\Ot(n)$. Merge the logarithmically many resulting functions
using Lemma~\ref{lem:merging}\textup{(i)} with total relative
error $O(\eps)$. The merge cost is $\Ot((1/\eps)^\rho)$,
and its additive error is $O(1/\eps)$ since the scaled optimum
is at most $2/\eps^2$.
The total cost is $\Ot(n+(1/\eps)^\beta)$ because
$\beta\ge\rho$.

The combined loss is $\Ot(1/\eps)$, or a relative
$\Ot(\eps)$ loss. Replacing the internal accuracy by a smaller
value within a polylogarithmic factor gives the requested
$1+\eps$ guarantee, as in the proof of Theorem~\ref{thm:main}.
All rounding is downward, and expanding the selected bundles
takes $O(n)$ time. This proves the reduction with witnesses.

\section{Finite precision in Chan's grid construction}\label{app:grids}
The grid family used in Lemma~\ref{lem:grids} is
\cite[Lemma~6]{Chan18}. We give its finite-mesh version to make
the bit complexity and failure probability explicit.
Assume $\delta$ and $\sigma$ are dyadic and
$0<\delta\le\sigma\le1/4$; constant-factor changes of accuracy
ensure these assumptions when the grid family is used.

Consider the targets in the $(\delta/16)$-grid of $[1,2]$.
For one such target $q$, the spacings whose multiples approximate
$q$ from below within $\delta/16$ form
\[
 [\sigma,2\sigma]\cap
 \bigcup_{k\ge1}
 \left[\frac{q-\delta/16}{k},\,\frac qk\right].
\]
As in Chan's proof, this union contains $\Theta(1/\sigma)$
disjoint intervals of length $\Theta(\delta\sigma)$, up to its
two boundary intervals. Its measure is $\Omega(\delta)$.
Choose a dyadic mesh on $[\sigma,2\sigma]$ of spacing at most
$\delta\sigma/1024$, within a factor of two of that bound.
Each interior interval contains a constant-factor proportion of
the mesh points predicted by its length.
A uniformly sampled mesh point is consequently good for $q$
with probability $\Omega(\delta/\sigma)$.

For failure budget $\zeta$, sample
\[
 O\left(\frac{\sigma}{\delta}
          \log\frac{1/\delta+2}{\zeta}\right)
\]
spacings. A union bound covers all $O(1/\delta)$ targets with
probability at least $1-\zeta$.
An arbitrary $p\in[1,2]$ is first rounded down to a target,
and then to a covered multiple, losing at most $\delta/8$.
In particular, every rounded profit is positive.

Enumerate and sort the multiples of the sampled spacings in
$[1-\delta/8,2]$ to verify coverage and assign input profit classes.
There are $\Ot(1/\delta)$ such multiples, including the logarithmic
dependence on $1/\zeta$.
Thus construction and verification cost $\Ot(1/\delta)$, and
assignment of $m$ classes costs $\Ot(m+1/\delta)$.
A failed verification is reported and may return the zero packing
function; it is included in the overall failure budget.
Since the mesh spacing is $\Omega(\delta^2)$, each spacing has
$O(\log(1/\delta))$ bits. The integer table indices and the
rounded profits also have short rational descriptions.

\section{Complementation and assembly of the head}\label{app:head}
This appendix records the adaptation of
\cite[Lemmas~6.12--6.15]{CLMZ25} used in the proof of
Lemma~\ref{lem:head}. The differences are the multiplicative
covering approximations supplied by Lemma~\ref{lem:prefixes} and
the corresponding enlarged cutoff.
Fix $\alpha<1/\eps^{1/2}$ and
$\delta=\Theta(\eps^{1/2}/\alpha)$.

\paragraph{Restricting the prefix and suffix}
Cap $\hat f_{I_L^j}$ at $1/\eps^{3/2}$.
Whenever $f_{I_L^j}(x)\le1/\eps^{3/2}$, this remains a lower
approximation with additive error at most
$O(\delta/\eps^{3/2})=O(1/(\alpha\eps))$.
For the prefix, complementation gives
\[
 f_{I_H^j}(x)
   =p(I_H^j)-h_{I_H^j}(w(I_H^j)-x).
\]
For an integer $x\ge0$, define
\begin{equation}\label{eq:high-approximation}
 F_H^j(x)=
 \begin{cases}
 p(I_H^j)-\hat h_{I_H^j}(w(I_H^j)-x),
  &\hat h_{I_H^j}(w(I_H^j)-x)\le(1+\delta)/\eps^{3/2},\\
 -\infty,&\text{otherwise}.
 \end{cases}
\end{equation}
A covering witness has weight at least $w(I_H^j)-x$ and profit
at most the reported covering loss. Its complement in $I_H^j$
therefore certifies $F_H^j(x)$.
If the exact covering loss is at most $1/\eps^{3/2}$, its upper
approximation is at most $(1+\delta)/\eps^{3/2}$.
Thus~\eqref{eq:high-approximation} retains every needed value,
with additive error at most $\delta/\eps^{3/2}$.
An exact cutoff at $1/\eps^{3/2}$ would not provide this guarantee.

\paragraph{The first merge}
The finite values of $F_H^j$ lie in
$[p(I_H^j)-(1+\delta)/\eps^{3/2},p(I_H^j)]$, and the capped
suffix values lie in $[0,1/\eps^{3/2}]$.
Subtract the two lower endpoints, apply
Lemma~\ref{lem:merging}\textup{(i)} with accuracy
$\Theta(\delta)$, and restore the translations.
The maximum translated value is $O(1/\eps^{3/2})$.
The resulting function $g_j$ is certified and adds error
$O(1/(\alpha\eps))$.
Its finite values lie in
$[p(I_H^j)-O(1/\eps^{3/2}),p(I_H^j)+1/\eps^{3/2}]$.
Negative reported profits are harmless lower bounds and are
replaced by zero when the final envelope is formed.

\paragraph{Adding the middle part}
Translate $g_j$ to a nonnegative range of length
$O(1/\eps^{3/2})$ and merge it with the common-profit functions
of $I_M^j$, as in the main proof.
Their maximum combined value is
$O(1/\eps^{3/2}+|I_M^j|/\eps)$.
The specified accuracy therefore adds only
$O(1/(\alpha\eps))$ error.
Restore the translation and call the result $q_j$.
All three underlying item sets are disjoint, so their witnesses
combine to a subset of the head.

\paragraph{The final envelope}
Every finite value of $q_j$ is at most $f_{I_{\mathrm{head}}}$.
Fix a capacity $x$, its interval $j$, and the optimum $S$ in
Lemma~\ref{lem:proximity}\textup{(ii)}.
Give each of the three parts the weight of its intersection with
$S$. Each restriction of $S$ is optimal for that part at its
assigned capacity, because a better restriction would improve $S$.
The prefix covering loss is $p(I_H^j\setminus S)$ and the suffix
profit is $p(S\cap I_L^j)$.
Both are at most $1/\eps^{3/2}$ and hence are retained, with
error $O(1/(\alpha\eps))$.
The two merges add only a constant number of errors of the same
size. Consequently,
\[
 q_j(x)\ge f_{I_{\mathrm{head}}}(x)-O(1/(\alpha\eps)).
\]
Taking $\max\{0,q_1,\ldots,q_q\}$ proves the claimed approximation
at every capacity.

\bibliographystyle{alphaurl}
\bibliography{references}

@article{AKMSW87,
  author = {Aggarwal, Alok and Klawe, Maria M. and Moran, Shlomo
            and Shor, Peter W. and Wilber, Robert E.},
  title = {Geometric Applications of a Matrix-Searching Algorithm},
  journal = {Algorithmica},
  volume = {2},
  pages = {195--208},
  year = {1987},
  doi = {10.1007/BF01840359}
}

@misc{AV26,
  author = {Alman, Josh and Vassilevska Williams, Virginia},
  title = {Truly Subquadratic {3SUM} and Truly Subcubic {APSP}
           via Triangles in Sparse Lopsided Graphs},
  year = {2026},
  howpublished = {arXiv:2610.06783},
  note = {Preprint, version 1, October 5, 2026},
  url = {https://arxiv.org/abs/2610.06783v1}
}

@inproceedings{AT19,
  author = {Axiotis, Kyriakos and Tzamos, Christos},
  title = {Capacitated Dynamic Programming: Faster Knapsack and Graph Algorithms},
  booktitle = {46th International Colloquium on Automata, Languages,
               and Programming (ICALP 2019)},
  series = {Leibniz International Proceedings in Informatics},
  volume = {132},
  pages = {19:1--19:13},
  publisher = {Schloss Dagstuhl--Leibniz-Zentrum f{\"u}r Informatik},
  year = {2019},
  doi = {10.4230/LIPIcs.ICALP.2019.19}
}

@inproceedings{Bri24,
  author = {Bringmann, Karl},
  title = {Knapsack with Small Items in Near-Quadratic Time},
  booktitle = {Proceedings of the 56th Annual ACM Symposium on Theory of Computing},
  pages = {259--270},
  publisher = {ACM},
  year = {2024},
  doi = {10.1145/3618260.3649719}
}

@inproceedings{BC22,
  author = {Bringmann, Karl and Cassis, Alejandro},
  title = {Faster Knapsack Algorithms via Bounded Monotone Min-Plus-Convolution},
  booktitle = {49th International Colloquium on Automata, Languages,
               and Programming (ICALP 2022)},
  series = {Leibniz International Proceedings in Informatics},
  volume = {229},
  pages = {31:1--31:21},
  publisher = {Schloss Dagstuhl--Leibniz-Zentrum f{\"u}r Informatik},
  year = {2022},
  doi = {10.4230/LIPIcs.ICALP.2022.31}
}

@inproceedings{Chan18,
  author = {Chan, Timothy M.},
  title = {Approximation Schemes for 0-1 Knapsack},
  booktitle = {1st Symposium on Simplicity in Algorithms (SOSA 2018)},
  series = {Open Access Series in Informatics},
  volume = {61},
  pages = {5:1--5:12},
  publisher = {Schloss Dagstuhl--Leibniz-Zentrum f{\"u}r Informatik},
  year = {2018},
  doi = {10.4230/OASIcs.SOSA.2018.5}
}

@misc{CIMY26,
  author = {Chan, Timothy M. and Itakura, Reina and Ma, Zhiyuan
            and Yu, Yuancheng},
  title = {{3SUM} Made Simpler (Without Matrix Multiplication),
           {$k$}-{SUM} Hypothesis Refuted for Odd {$k$},
           and Real {3SUM}/{APSP} Derandomized},
  year = {2026},
  howpublished = {arXiv:2610.42012},
  note = {Preprint, version 2, November 14, 2026},
  url = {https://arxiv.org/abs/2610.42012v2}
}

@inproceedings{CLMZ24,
  author = {Chen, Lin and Lian, Jiayi and Mao, Yuchen and Zhang, Guochuan},
  title = {Faster Algorithms for Bounded Knapsack and Bounded Subset Sum
           via Fine-Grained Proximity Results},
  booktitle = {Proceedings of the 2024 Annual ACM-SIAM Symposium
               on Discrete Algorithms},
  pages = {4828--4848},
  publisher = {SIAM},
  year = {2024},
  doi = {10.1137/1.9781611977912.171}
}

@article{CLMZ25,
  author = {Chen, Lin and Lian, Jiayi and Mao, Yuchen and Zhang, Guochuan},
  title = {A Nearly Quadratic-Time {FPTAS} for Knapsack},
  journal = {SIAM Journal on Computing},
  pages = {STOC24-1--STOC24-23},
  year = {2025},
  doi = {10.1137/24M1686772}
}

@inproceedings{CLMZ25weak,
  author = {Chen, Lin and Lian, Jiayi and Mao, Yuchen and Zhang, Guochuan},
  title = {Weakly Approximating Knapsack in Subquadratic Time},
  booktitle = {52nd International Colloquium on Automata, Languages,
               and Programming (ICALP 2025)},
  series = {Leibniz International Proceedings in Informatics},
  volume = {334},
  pages = {51:1--51:18},
  publisher = {Schloss Dagstuhl--Leibniz-Zentrum f{\"u}r Informatik},
  year = {2025},
  doi = {10.4230/LIPIcs.ICALP.2025.51}
}

@inproceedings{CDXZ22,
  author = {Chi, Shucheng and Duan, Ran and Xie, Tianle and Zhang, Tianyi},
  title = {Faster Min-Plus Product for Monotone Instances},
  booktitle = {Proceedings of the 54th Annual ACM SIGACT Symposium
               on Theory of Computing},
  pages = {1529--1542},
  publisher = {ACM},
  year = {2022},
  doi = {10.1145/3519935.3520057}
}

@article{CMWW19,
  author = {Cygan, Marek and Mucha, Marcin and W{\k{e}}grzycki, Karol
            and W{\l}odarczyk, Micha{\l}},
  title = {On Problems Equivalent to {$(\min,+)$}-Convolution},
  journal = {ACM Transactions on Algorithms},
  volume = {15},
  number = {1},
  pages = {14:1--14:25},
  year = {2019},
  doi = {10.1145/3293465}
}

@inproceedings{DJM23,
  author = {Deng, Mingyang and Jin, Ce and Mao, Xiao},
  title = {Approximating Knapsack and Partition via Dense Subset Sums},
  booktitle = {Proceedings of the 2023 Annual ACM-SIAM Symposium
               on Discrete Algorithms},
  pages = {2961--2979},
  publisher = {SIAM},
  year = {2023},
  doi = {10.1137/1.9781611977554.ch113}
}

@article{IK75,
  author = {Ibarra, Oscar H. and Kim, Chul E.},
  title = {Fast Approximation Algorithms for the Knapsack and Sum of Subset Problems},
  journal = {Journal of the ACM},
  volume = {22},
  number = {4},
  pages = {463--468},
  year = {1975},
  doi = {10.1145/321906.321909}
}

@inproceedings{Jin19,
  author = {Jin, Ce},
  title = {An Improved {FPTAS} for 0-1 Knapsack},
  booktitle = {46th International Colloquium on Automata, Languages,
               and Programming (ICALP 2019)},
  series = {Leibniz International Proceedings in Informatics},
  volume = {132},
  pages = {76:1--76:14},
  publisher = {Schloss Dagstuhl--Leibniz-Zentrum f{\"u}r Informatik},
  year = {2019},
  doi = {10.4230/LIPIcs.ICALP.2019.76}
}

@inproceedings{Jin24,
  author = {Jin, Ce},
  title = {0-1 Knapsack in Nearly Quadratic Time},
  booktitle = {Proceedings of the 56th Annual ACM Symposium on Theory of Computing},
  pages = {271--282},
  publisher = {ACM},
  year = {2024},
  doi = {10.1145/3618260.3649618}
}

@inproceedings{JPSX26,
  author = {Jin, Ce and Park, Jaewoo and Saha, Barna and Xu, Yinzhan},
  title = {Deterministic Monotone Min-Plus Product and Convolution},
  booktitle = {53rd International Colloquium on Automata, Languages,
               and Programming (ICALP 2026)},
  series = {Leibniz International Proceedings in Informatics},
  volume = {374},
  pages = {119:1--119:23},
  publisher = {Schloss Dagstuhl--Leibniz-Zentrum f{\"u}r Informatik},
  year = {2026},
  doi = {10.4230/LIPIcs.ICALP.2026.119}
}

@article{KP04,
  author = {Kellerer, Hans and Pferschy, Ulrich},
  title = {Improved Dynamic Programming in Connection with an {FPTAS}
           for the Knapsack Problem},
  journal = {Journal of Combinatorial Optimization},
  volume = {8},
  number = {1},
  pages = {5--11},
  year = {2004},
  doi = {10.1023/B:JOCO.0000021934.29833.6b}
}

@book{KPP04,
  author = {Kellerer, Hans and Pferschy, Ulrich and Pisinger, David},
  title = {Knapsack Problems},
  publisher = {Springer},
  address = {Berlin},
  year = {2004},
  doi = {10.1007/978-3-540-24777-7}
}

@inproceedings{KPS17,
  author = {K{\"u}nnemann, Marvin and Paturi, Ramamohan and Schneider, Stefan},
  title = {On the Fine-Grained Complexity of One-Dimensional Dynamic Programming},
  booktitle = {44th International Colloquium on Automata, Languages,
               and Programming (ICALP 2017)},
  series = {Leibniz International Proceedings in Informatics},
  volume = {80},
  pages = {21:1--21:15},
  publisher = {Schloss Dagstuhl--Leibniz-Zentrum f{\"u}r Informatik},
  year = {2017},
  doi = {10.4230/LIPIcs.ICALP.2017.21}
}

@article{Law79,
  author = {Lawler, Eugene L.},
  title = {Fast Approximation Algorithms for Knapsack Problems},
  journal = {Mathematics of Operations Research},
  volume = {4},
  number = {4},
  pages = {339--356},
  year = {1979},
  doi = {10.1287/moor.4.4.339}
}

@inproceedings{Mao24,
  author = {Mao, Xiao},
  title = {{$(1-\epsilon)$}-Approximation of Knapsack in Nearly Quadratic Time},
  booktitle = {Proceedings of the 56th Annual ACM Symposium on Theory of Computing},
  pages = {295--306},
  publisher = {ACM},
  year = {2024},
  doi = {10.1145/3618260.3649677}
}

@misc{Wang26,
  author = {Wang, Jionghan},
  title = {A Faster Deterministic {3SUM} Solver and Its Weighted-Cache
           Optimization Limit},
  year = {2026},
  howpublished = {Lean 4 formal development},
  note = {Revision e1ae71db2231; theorem \texttt{Race.bound}
          gives the exponent 1.920279408185},
  url = {https://github.com/Wang-jiong-han/3sum-weighted-cache/tree/e1ae71db2231f172c6541820e2ce0a54aa14fc64}
}

@misc{Zhou26,
  author = {Zhou, Huanyi},
  title = {Bounded Knapsack in Truly Subquadratic Time},
  year = {2026},
  howpublished = {arXiv:2611.15792},
  note = {Preprint, version 1, November 11, 2026},
  url = {https://arxiv.org/abs/2611.15792v1}
}
\end{document}